\documentclass[aps,12pt,showpacs,preprintnumbers,nofootinbib]{revtex4}

\usepackage{graphicx}
\usepackage{amsmath,amssymb}
\newcommand\+{\dagger}
\renewcommand\d{\partial}
\newcommand\x{{\bf x}}
\newcommand\p{{\bf p}}
\newcommand\q{{\bf q}}
\renewcommand\k{{\bf k}}
\newcommand\<{\langle}
\renewcommand\>{\rangle}
\renewcommand\Im{\mathop{\mathrm{Im}}}
\begin{document}

\preprint{INT-PUB-10-009}
\title{Short-distance and short-time structure of a unitary Fermi gas}
\author{Dam~T.~Son and Ethan~G.~Thompson}
\affiliation{Institute for Nuclear Theory, University of Washington,
Seattle, WA 98195-1550, USA}

\begin{abstract}
We consider the operator product expansions for unitarity fermions.
We compute the dynamic structure factor $S(\q,\omega)$ at large
frequency and wavenumber away from the
one-particle peak.  The overall normalization of $S(\q,\omega)$
is determined by Tan's contact parameter, and the dependence on 
$\q$ and $\omega$ is obtained in closed analytic form.  We also find
energy deposited into the system by small, rapid variations
of the inverse scattering length.
\end{abstract}

\maketitle

\section{Introduction}

One of the most interesting systems currently under study 
is that of fermions with
interactions fine-tuned to unitarity~\cite{Leggett}.  
This system is interesting due
to the absence of any scale parameter other than the density.  The
system is strongly interacting and thus presents a challenge for many-body
methods.

Recently, these systems have been probed with high-frequency or
short-wavelength external probes.  By bombarding the fermions with
radio-frequency (RF) radiation, which causes a transition from one
atomic state to another, one can probe the structure of the unitary
Fermi gas.  Previous investigations~\cite{Strinati,Schneider,Zwerger} 
have shown that the high frequency tail of the RF line shape is related to the
contact parameter, first introduced by Shina
Tan~\cite{Shina1,Shina2}.

In this paper, we address two problems.  The first one concerns the
dynamic structure factor $S(\q,\omega)$.  This factor, in principle,
can be measured by Bragg scatterings~\cite{Drummond}.
At the Fermi momentum and energy scales, the dynamic structure function
depends on the complicated many-body physics of the unitary Fermi gas.
For much larger $\q$ and $\omega$, the picture becomes simpler.
There is a dominant coherent peak in the
response function at $\omega=\epsilon_\q$.  
Far away from this peak, the strength of the
dynamic structure factor should be proportional to Tan's parameter, 
the same parameter that characterizes the RF tail.  It is easy to 
understand why this is true:
in order to absorb an external ``virtual photon''
carrying large energy and momentum and far away from the coherent peak,
the photon has to hit a particle when it is near a second particle,
so that it can give its momentum and energy to both.  The absorption
rate, therefore, depends on the probability of finding two particles
at short distance from each other---which is characterized by Tan's
contact parameter.  

The problem of the dynamic structure factor is more complicated than that
of the RF response.  In particular, care should be taken to not
violate conservation laws.  In this paper we use the operator product
expansion (OPE) to facilitate the calculation.  The operator product expansion,
introduced by Wilson, is a standard method of quantum field theory.
The use of the OPE for cold atoms was pioneered by Braaten and
Platter~\cite{Braaten:2008uh}.  The end result for the dynamic
structure factor is Eq.~(\ref{DSF}). 

The second problem considered in this paper is the 
calculation of the energy deposition into a unitary
Fermi gas by small, rapid variations of the inverse scattering length.
It can be solved using the same OPE methods.

The usefulness of the operator product expansion can be illustrated as
follows.  Suppose we need to compute the following Green's function
\begin{equation}
  G_{AB}(\omega,\q) =
  \int\! dt\, d\x\, e^{i\omega t - i\q\cdot\x}\, \< A(t,\x) B(0,{\bf 0})\>
\end{equation}
for large $\omega$ and $\q$.  Here ``large'' means energy and momentum
 much larger than the typical energy and
momentum scales of the state with respect to which the average
$\<\ldots\>$ is taken.  For the ground state of a unitary Fermi gas,
these scales are the Fermi energy and Fermi momentum.  Let us also
recall that one can associate a local operator $O$ with a scaling
dimension $\Delta_O$.  In our counting scheme, the dimension of momentum is
1 and of energy is 2 (the particle mass $m$ is set to 1). 
Assuming that the operator product expansion
exists, one can expand the product $A(t,\x)B(0,{\bf 0})$ in terms of
local operators,
\begin{equation}
  A(t,\x) B(0,{\bf 0}) = \sum_i |\x|^{\Delta_i-\Delta_A-\Delta_B}
   f_i \left( \frac{|\x|^2}t \right) O_i(0).
\end{equation}
Here $f_i$ are functions of one variable $|\x|^2/t$,
and $\Delta_i$ are the dimensions of $O_i$.  In contrast to the
OPE in relativistically invariant theories, in nonrelativistic
theories the OPE coefficients are not constant, but are in general
functions of this variable.  This identity is to be interpreted as an
operator identity; in particular, we can take the expectation value of
both sides with respect to any state, including thermodynamic states.
Taking the average and performing a Fourier transform, one finds
\begin{equation}
  G_{AB}(\omega, \q) = \sum_i 
  \frac1{\omega^{(5+\Delta_i-\Delta_A-\Delta_B)/2}}
  c_i \left( \frac{q^2}\omega\right) \<O_i\>, 
  \qquad q \equiv |\q|.
\end{equation}
On the right hand side, the higher the dimension of $O_i$, the more
rapidly its contribution decays in the limit of large momentum/energy
(to be precise, the limit considered in this paper will be $\omega\to\infty$, $\q\to\infty$, $q^2/\omega=\textrm{fixed}$.)  Thus,
the leading behavior of the Green function is dominated by those few 
operators in the OPE with smallest scaling dimension.

The expectation values of the operators $O_i$, are, in general, not
computable theoretically because they depend on many-body physics.
Thus, they should be considered as numbers parameterizing the many-body
state.  The OPE coefficients $c_i$, however, depend only on few-body
physics (although, the number of bodies increases with increasing
complexity of the operator $O_i$), and hence can be computed reliably,
at least for simple operators $A$, $B$ and $O_i$.  In this way, the
functional dependence of $G_{AB}$ on frequency and wavenumber can be
expressed in terms of a few numbers which have to be determined
experimentally or numerically.

We now discuss the operators of lowest dimension.  The unitary fermions are
described by the Lagrangian
\begin{equation}
  {\cal L} = i\psi^\+ \d_t \psi - \frac{|\nabla\psi|^2}{2} 
    + \psi_2 \psi_1\phi^*
    + \psi_1^\+ \psi_2^\+ \phi
    - c_0^{-1} \phi^*\phi.
\end{equation}
We set the fermion mass to one.  The indices $1$, $2$ refer to the two
spin polarizations.  If one integrates out $\phi$, the saddle point
for this field is at $\phi=c_0\psi_2\psi_1$.  This Lagrangian is
therefore equivalent to that with a four-Fermi interaction
$c_0\psi_1^\+\psi_2^\+\psi_2\psi_1$.  We will use dimensional
regularization, where setting $c_0^{-1}=0$ corresponds to fine-tuning
the interaction to infinite scattering length.  The propagator of the
$\phi$ field is completely determined by its self-energy, and is equal
to
\begin{equation}\label{scalar-prop}
  D(\omega,\q) = -\frac{4\pi}{\sqrt{q^2/4-\omega-i0}}\,.
\end{equation}

Let us discuss the operators which may have nonzero expectation value
in a unitary Fermi gas.  For our applications, we will need to
consider only operators which do not carry particle numbers.
Moreover, we assume the ground state to be isotropic, so we need to
look only at operators with vanishing orbital angular momentum.  The following
two operators have the lowest dimensions: $n_i=\psi_i^\+\psi_i$ ($i=1,2$) and
$\phi^*\phi$.  The operators $n_i$ are the particle number densities and
have dimension 3.  The operator $\phi^*\phi$ has dimension equal to 4.  This
can be seen by computing the dimension of $\phi$ from
Eq.~(\ref{scalar-prop}).  An alternative way to find the dimension of
$\phi$ is to use the operator-state correspondence, according to which
the dimension of $\phi$ is the ground state energy of a system of one
spin-up and one spin-down particle in a harmonic trap with unit
oscillator frequency.  This ground state energy is 2.

As we shall see, the expectation value $\<\phi^*\phi\>$ can be
identified with Tan's contact parameter.  Together, these two
operators dominate the high-momentum behavior of correlation
functions, including the dynamic structure factor.

\section{Single-particle Green's function}
To establish the relationship between $\<\phi^*\phi\>$ and Tan's
contact parameter, let us apply the method of the 
OPE to the one-particle Green's function
\begin{equation}
  iG(\omega,\q) = \int\!dt\,d\x\, e^{i\omega t- i\q\cdot\x}\,
  \< T  \psi_1(t,{\bf x}) \psi^\dagger_1(0,{\bf 0}) \>,
\end{equation}
where $T$ denotes time ordering.\footnote{Y.~Nishida~\cite{Nishida}
has essentially arrived at a similar derivation.}  
At the end we will be interested only
in the case of $t<0$, where the Green's function is
 $-\<\psi^\+(0)\psi(t,\x)\>$,
but the Feynman diagrams are readily available for the time-ordered
Green's function.

According to the previous discussion, we can write
\begin{equation}\label{fermionprop}
  G(\omega,\q) = C_n(\omega,\q)\<\psi_2^\+\psi_2\> 
   + C_{\phi^*\phi}(\omega,\q) \<\phi^*\phi\> +\cdots
\end{equation}
(We shall see why there is no $\<\psi_1^\+\psi_1\>$ term below.)

\begin{figure}[ht]
\includegraphics[width=0.25\textwidth]{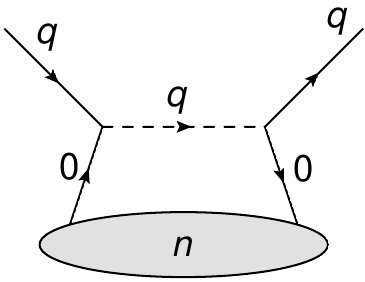}
\caption{The diagram that gives $C_n$ in Eq.~(\ref{fermionprop}).  
The large momentum and energy flow along lines carrying label ``q.''}
\label{fig:Cn_fermionprop}
\end{figure}

To compute $C_n$, we use the Feynman diagram in
Fig.~\ref{fig:Cn_fermionprop}.

Physically, this diagram describes the interaction of a particle
carrying large momentum and frequency $(\omega,\q)$ with particles already
in the medium.  The energy and momentum of the particles in the medium
can be neglected compared to $(\omega,\q)$.  The ``hard'' particle
interacts with particles of type 2 in the medium with the result,
\begin{equation}
  C_n(\omega,\q) = -\frac{4\pi}{(\omega-q^2/2+i0)^2
  \sqrt{q^2/4-\omega-i0}}\,.
\end{equation}
One can see immediately that $C_n(\omega,\q)$ has singularities only
in the lower-half plane, and thus does not contribute to the Green's
function for $t<0$.  This can already be seen from the Feynman diagram in
Fig.~\ref{fig:Cn_fermionprop}: one can go directly from the 
initial point to the
final point, following the direction of the propagators (fermions and
scalars).  Recall that the propagators are all retarded.

\begin{figure}[ht]
\includegraphics[width=0.4\textwidth]{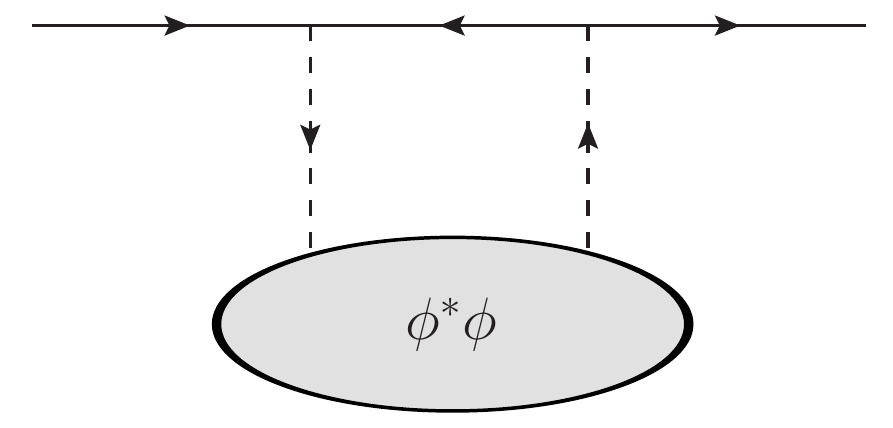}
\caption{A contribution of $\<\phi^*\phi\>$ to the fermion propagator}
\label{fig:ferm-prop}
\end{figure}

\begin{figure}[ht]
\includegraphics[width=0.75\textwidth]{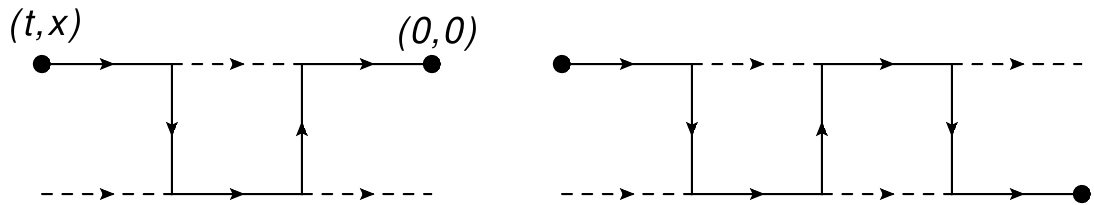}
\caption{The other diagrams contributing to the fermion propagator}
\label{fig:ferm-prop-others}
\end{figure}

Consider now the OPE coefficient of $C_{\phi^*\phi}$. It
 can be found by 
computing the fermion propagator in the background field of $\phi$.
One diagram that contributes to $C_{\phi^*\phi}$ 
is as in Fig.~\ref{fig:ferm-prop}.
The dashed lines carry zero energy and momentum, and go to the
``condensate'' $\<\phi^*\phi\>$.  Fig.~\ref{fig:ferm-prop} is not
the sole diagram contributing to $C_{\phi^*\phi}$;
there is an infinite number of diagrams contributing to this coefficient,
like those depicted in
Fig.~\ref{fig:ferm-prop-others} and obvious subsequent iterations.  
It can be seen that, in all diagrams of the type of 
Fig.~\ref{fig:ferm-prop-others}, one can go from one end of the diagram to the 
other by
following the arrows, which is not true in the case of 
Fig.~\ref{fig:ferm-prop}.  Thus, if our ultimate goal is to compute the
Green's function for $t<0$, it is sufficient to just evaluate the diagram in Fig.~\ref{fig:ferm-prop}.

From the Feynman rules, the contribution of Fig.~\ref{fig:ferm-prop} to
$C_{\phi^*\phi}$ is 
\begin{equation}
  -\frac 1{(\omega -\epsilon_\q+i0)^2(-\omega-\epsilon_\q+i0)} ,
\end{equation}
where $\epsilon_\q=q^2/2$.

With this information we can find the leading nontrivial contribution
to the Green's function for $t<0$:
\begin{multline}
  \int\!{d\x}\, e^{-i\q\cdot\x}\<\psi_1^\dagger(0,{\bf 0})\psi_1 (t,\x)\> =
  -i\int\!\frac{d\omega}{2\pi}\, e^{-i\omega t}
  C_{\phi^*\phi}(\omega,\q) \<\phi^*\phi\>\\
  = \frac1{q^4}\exp \left(i\frac{q^2}2 t\right) \<\phi^*\phi \>
    + \cdots , \qquad t<0.
\end{multline}
In the limit $t\to-0$, we find the tail of the distribution function to be
\begin{equation}
  n_\q = \frac{\<\phi^*\phi\>}{q^4},
\end{equation}
which allows us to establish $\<\phi^*\phi\> = C$, where 
$C$ is Tan's contact parameter~\cite{Shina1,Shina2}.

One can also show that the diagrams in
Fig.~\ref{fig:ferm-prop-others} do not contribute to the imaginary
part of $G(\omega,\q)$ for $\omega<0$.  The latter receives only a 
contribution from Fig.~\ref{fig:ferm-prop}, which is
\begin{equation}
  \omega < 0: \qquad
  \Im G(\omega,\q) = \frac{\<\phi^*\phi\>}{q^4}\pi\delta(\omega+\epsilon_\q).
\end{equation}
The peak at $\omega=-\epsilon_\q$ is 
already discussed in Refs.~\cite{Combescot,Randeria-bendback}.  At
this level, we are not able to resolve the structure of the peak.

\section{RF spectroscopy}
\begin{figure}[h]
\includegraphics[width=0.45\textwidth]{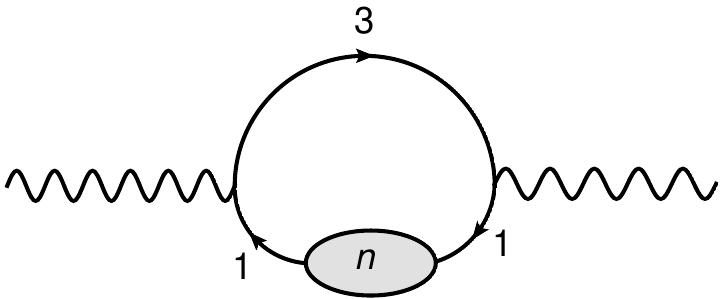}
\caption{Diagram contributing to $C_n$}
\label{fig:n-O13}
\end{figure}

Consider now a system where, in addition to the `up' and `down' fermions 
included in the earlier Lagrangian, a third species of fermion is added 
which does not interact with the other two fermions.  Suppose we turn on 
a photon field which converts atoms of type one into
atoms of type three. The absorption rate is proportional to the imaginary 
part of the Green's function of
\begin{equation}
  O_{13} = \psi_3^\+\psi_1.
\end{equation}
We thus need to compute the OPE expansion of 
\begin{equation}
 \Pi(\omega, \q) = -i \int\!dt\,d\x\, 
  e^{i\omega t - i\q\cdot\x}\, \<T O_{13}(t,\x) O_{13}^\+(0,{\bf 0})\>
  = C_n \<n_1\> + C_{\phi^*\phi}\<\phi^*\phi\> + \cdots
\end{equation}
The coefficient $C_n$ is obtained from the diagram of Fig.~\ref{fig:n-O13},
\begin{equation}\label{n-coeff}
  C_n = \frac 1{\omega - \epsilon_\q + i0} .
\end{equation}
The external wavy lines in this diagram represent the operator $O_{13}$. The interpretation of this formula is obvious.  The dominant part of
the response function is the same as a collection of noninteracting
particles with zero momentum, and has a peak at $\omega=\epsilon_\q$.  

\begin{figure}[t]
\includegraphics[width=0.50\textwidth]{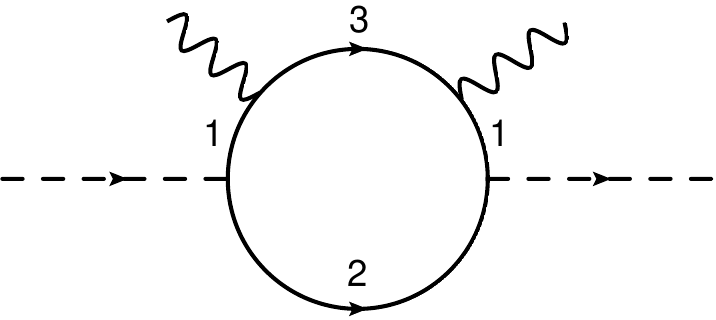}
\caption{Diagram contributing to $C_{\phi^*\phi}$.  
The external dashed lines, which
carry zero energy and momentum,
go to the ``condensate'' $\<\phi^*\phi\>$ which is not depicted for 
simplicity.}
\label{fig:phiphi-O13}
\end{figure}

Let us now compute the diagram which determines the coefficient of
$\phi^*\phi$ in the OPE (Fig.~\ref{fig:phiphi-O13}).
We find
\begin{equation}
\begin{split}
  \Pi(\omega,\q) &= \cdots -i \int\! \frac{dp_0\,d\p}{(2\pi)^4}\,
  \frac1{(p_0-\epsilon_\p+i0)^2}\,\frac1{p_0 + \omega - \epsilon_{\p+\q}}\,
  \frac1{p_0+\epsilon_\p-i0} \< \phi^*\phi\>\\ 
  & \qquad\qquad\qquad + (\omega\to-\omega,\q\to-\q)
  + \cdots\\
  &= \cdots+\int\!\frac{d\p}{(2\pi)^3}\, \frac1{4\epsilon_\p^2}\,
   \frac1{\omega-\epsilon_p-\epsilon_{\p+\q}+i0} \<\phi^*\phi\> 
  + (\omega\to-\omega,\q\to-\q)
  +\cdots
\end{split}
\end{equation}
This integral is divergent in the infrared.  We can isolate the divergent
part of the integral as
\begin{equation}
  \frac{\<\phi^*\phi\>}{\omega - \epsilon_\q+i0}
  \int\!\frac{d\p}{(2\pi)^3}\, \frac1{4\epsilon_\p^2}\, 
  + \int\!\frac{d\p}{(2\pi)^3}\, \frac1{4\epsilon_\p^2}\left(
   \frac1{\omega-\epsilon_\p-\epsilon_{\p+\q}+i0} -
   \frac1{\omega-\epsilon_\q+i0} \right) \<\phi^*\phi\>.
\end{equation}

It is clear that the first (divergent) part is actually included in the
first term in the OPE (\ref{n-coeff}), i.e., it represents the contribution of the
condensate $\<\phi^*\phi\>$ to the number density $n$.  Subtracting
this contribution, the second part gives $C_{\phi^*\phi}$,
\begin{equation}
  C_{\phi^*\phi} = \int\!\frac{d\p}{(2\pi)^3}\, \frac1{4\epsilon_p^2}\left(
   \frac1{\omega-\epsilon_p-\epsilon_{\p+\q}+i0} -
   \frac1{\omega-\epsilon_\q+i0} \right)
  + (\omega\to-\omega, \q\to-\q).
\end{equation}

We  now restrict ourselves to the imaginary part at $\q=0$.  We have
\begin{equation}
  I(\omega) = -\frac1\pi \Im  \Pi(\omega, {\bf 0}) = 
  \frac{\<\phi^*\phi\>}{4\pi^2} \frac1{\omega^{3/2}}\,,
\end{equation}
which coincides with the result obtained previously by other
authors~\cite{Schneider,Zwerger}.

\section{Dynamic structure factor}

Having tested the OPE method on the simple examples above, we will now
discuss the dynamic structure factor.  The dynamic structure factor can be
defined as the imaginary part of the Green function of the density,
\begin{equation}
  G_{nn} (\omega,\q) = -i \int\!dt\,d\q\, e^{i\omega t- i\k\cdot\x}\,
  \< T n(t,\x) n(0,{\bf 0})\> = C_n \<n\> + C_{\phi^*\phi} \< \phi^*\phi\>
  + \cdots
\end{equation}
Namely,
\begin{equation}
  S(\q,\omega) = -\frac1\pi \Im G_{nn}(\omega,\q), \qquad
  \omega > 0.
\end{equation}

The coefficient $C_n$ is
\begin{equation}
  C_n = G(\omega,\q) + G(-\omega,-\q) = 
  \frac{2\epsilon_\q}{\omega^2-\epsilon_\q^2+i0}\,,
\end{equation}
and corresponds to the one-particle peak $\sim\delta(\omega-\epsilon_\q)$
in the dynamic structure factor.  This peak dominates all sum rules.  However,
we are interested in the structure factor far away from this peak, hence
we need to compute the OPE coefficient of the next operator, $\phi^*\phi$. 

\begin{figure}[ht]
\includegraphics[width=0.75\textwidth]{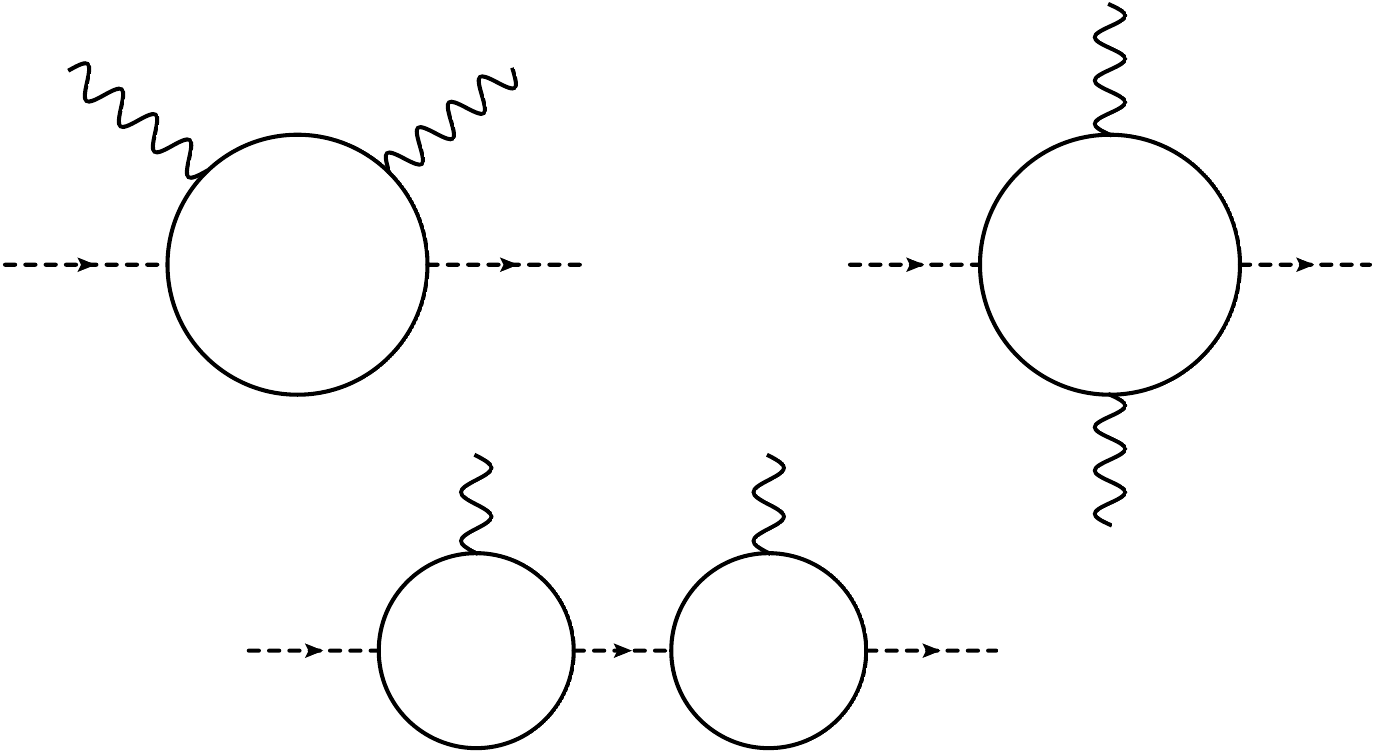}
\caption{Schematic diagrams contributing to $C_{\phi\phi}$. 
Each diagram represents a number of Feynman graphs where the photon lines
are attached to different fermion lines}
\label{fig:phiphi-nn}
\end{figure}

The diagrams contributing to $C_{\phi^*\phi}$ are sketched
schematically in Fig.~\ref{fig:phiphi-nn}.

The contribution from the first diagram is exactly twice the diagram
in the RF case,
\begin{equation}
  C^{(1)}_{\phi^*\phi} = 2 \int\!\frac{d\p}{(2\pi)^3}\, 
   \frac1{4\epsilon_\p^2}\left(
   \frac1{\omega-\epsilon_\p-\epsilon_{\p+\q}+i0} -
   \frac1{\omega-\epsilon_\q+i0} \right)
  + (\omega\to-\omega,\q\to-\q).
\end{equation}
The second diagram gives
\begin{equation}
  C^{(2)}_{\phi^*\phi} = 2 \int\!\frac{d\p}{(2\pi)^3}\, 
  \frac1{4\epsilon_\p\epsilon_{\p+\q}}\left(
  \frac1{\omega-\epsilon_\p-\epsilon_{\p+\q}+i0}-
  \frac1{\omega+\epsilon_\p+\epsilon_{\p+\q}-i0}\right).
\end{equation}
The third diagram is the product of three pieces, each of which can
be computed separately.  At the end one finds
\begin{multline}
  C^{(3)}_{\phi^*\phi} = \left[ -2 \int\!\frac{d\p}{(2\pi)^3}\,
  \frac1{2\epsilon_\p} \frac1{\omega-\epsilon_\p - \epsilon_{\p+\q}+i0}
  \right]^2  \frac{-4\pi}{\sqrt{\frac12\epsilon_\q-\omega-i0}}
  + (\omega\to-\omega, \q\to-\q).
\end{multline}
Evaluating the imaginary part of the correlation function
for positive $\omega$, we find
that it is zero when $\omega<q^2/4$.  This should be the case since
$\omega=q^2/4$ is the threshold for knocking out two particles from the
medium (recall that all the energy scales associated with the medium, 
like the chemical potential, are negligible).
For $\omega>q^2/4$, the dynamic structure factor is
\begin{multline}\label{DSF}
 \frac{S(\q,\omega)}{\<\phi^*\phi\>}
  = -\frac1\pi\Im  C_{\phi^*\phi} = \frac1{2\pi^2}
  \frac{\sqrt{\omega-q^2/4}}{(\omega-q^2/2)^2} 
  + \frac1{2\pi^2\omega q} \ln \frac{\omega+q\sqrt{\omega-q^2/4}}
    {|\omega-q^2/2|}\\
  - \frac1{\pi^2 q^2\sqrt{\omega-q^2/4}} \left(
  \ln^2 \frac{\omega+q\sqrt{\omega-q^2/4}}{|\omega-q^2/2|}
  -\pi^2\theta(q^2/2-\omega)\right).
\end{multline}

\begin{figure}[ht]
\includegraphics[width=0.5\textwidth]{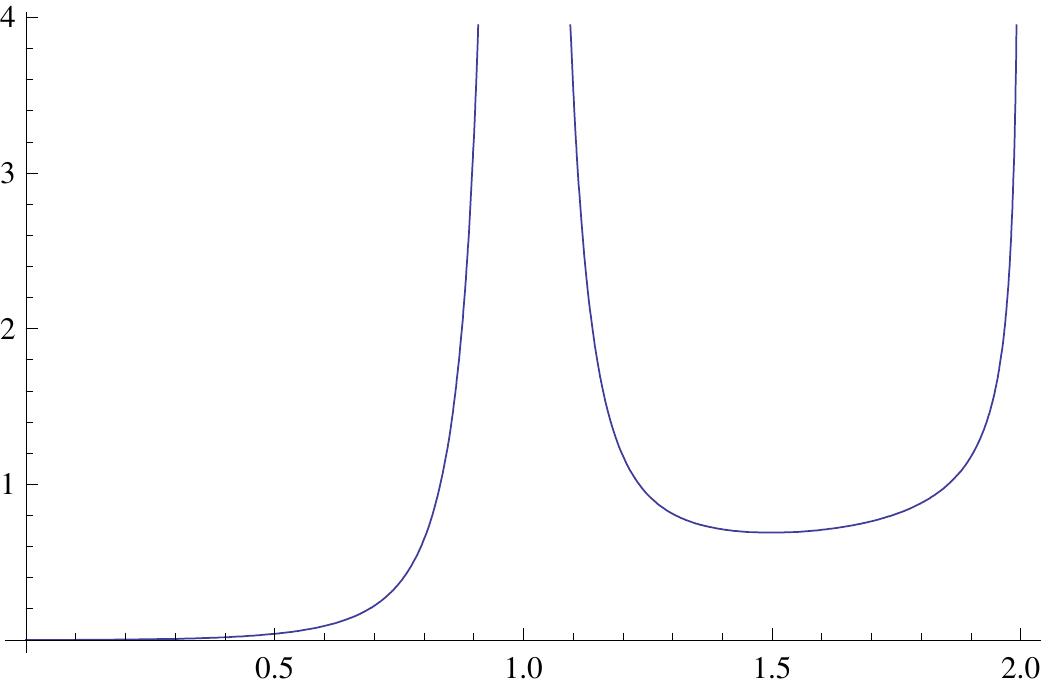}
\caption{A plot of $\omega^{3/2}S(\q,\omega)/\<\phi^*\phi\>$ as a
function of $x=q^2/2\omega$.}
\label{fig:fixom}
\end{figure}

\begin{figure}[ht]
\includegraphics[width=0.5\textwidth]{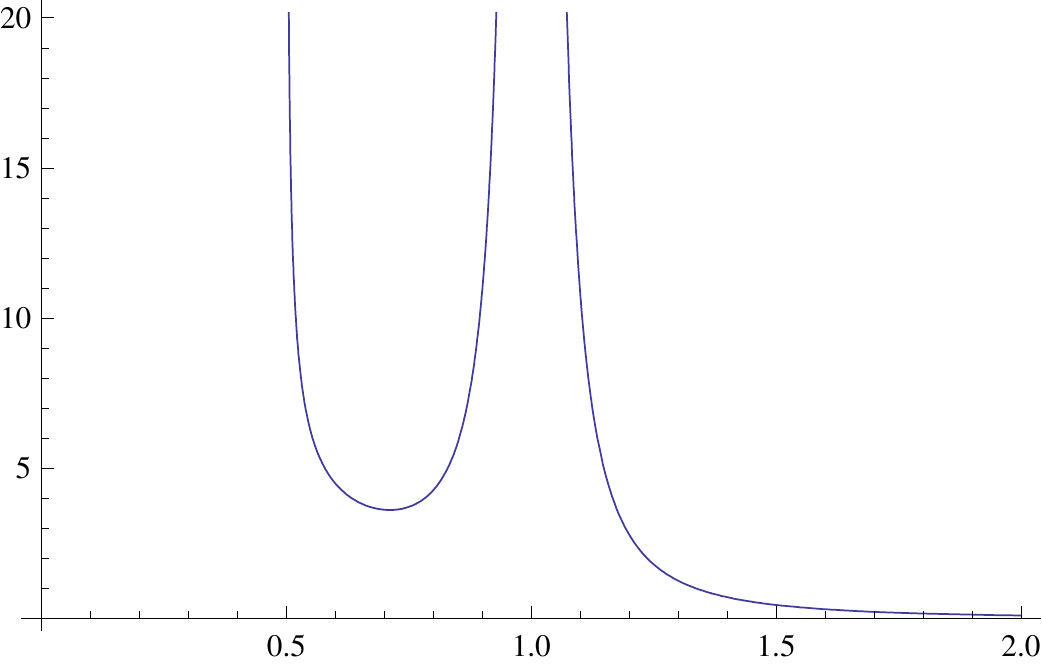}
\caption{A plot of $q^{3/2}S(\q,\omega)/\<\phi^*\phi\>$ as a
function of $x^{-1}=2\omega/q^2$.}
\label{fig:fixk}
\end{figure}

In Fig.~\ref{fig:fixom} we plot the function
$\omega^{3/2}S(\omega,\q)/\<\phi^*\phi\>$ as a
function of the ``Bjorken $x$'' variable, $x=q^2/2\omega$.  This
can be thought of as a plot of $S$ at fixed $\omega$ as a function of $q^2$.
In Fig.~\ref{fig:fixk}, 
we plot the same quantity as a function
of $x^{-1}=2\omega/q^2$.
This plot is basically a plot of $S$ as a function of $\omega$ at fixed $q$.

As we see from the plots, the dynamic structure factor is positive for
all $\omega$ and $\q$, as it should be.  It vanishes in the zero
momentum, finite frequency limit $\q=0$, as it should due to the
conservation of particle number.  At small $q^2/\omega$, the behavior
of the dynamic structure factor is
\begin{equation}\label{S-smallq}
  S(\q,\omega) = \frac4{45\pi^2} \frac{q^4}{\omega^{7/2}}\<\phi^*\phi\>.
\end{equation}
The dynamic structure function has a singularity near $\omega=q^2/2$,
which is the single-particle peak.  It would be incorrect to take 
the limit $x= q^2/2\omega\to1$ in our formula at fixed $\q$ or fixed
$\omega$.  Our result is strictly valid in
the regime of $\omega\to\infty$, $\q\to\infty$, fixed $x=q^2/2\omega\neq1$.
The weaker singularity at $\omega=q^2/4$ ($x=2$) is associated with the fact
that near this point, the two particles that are kicked out have small
relative momentum, and thus interact strongly with each other.

\section{Response of unitary gas to rapid changes of the inverse
scattering length}

We consider in this section the following problem.  Assume we make the
inverse scattering length $\alpha\equiv a^{-1}$ exhibit periodic
behavior in time,
\begin{equation}
  \alpha(t) = a^{-1}(t) = \alpha_0 \cos\omega t,
\end{equation}
and assume the amplitude $\alpha_0$ to be small, and the frequency
of the oscillations to be large compared to the Fermi energy, 
$\omega\gg\epsilon_{\rm F}$.  The
question is: at what rate is energy deposited into the system?

We know that the inverse scattering length is coupled to the operator
$\phi^*\phi$, i.e., introducing a finite inverse scattering length
corresponds to adding into the Lagrangian a term
\begin{equation}
  {\cal L} \to {\cal L} + \frac{\alpha}{4\pi} \phi^*\phi.
\end{equation}

The energy deposited into the system, in unit volume and in unit time,
can be computed from the formula
\begin{equation}
  \frac{d\epsilon}{dt} = \frac{\alpha^2_0}{2(4\pi)^2} \omega \,
  |\Im G_{OO}(\omega,{\bf 0})|,
\end{equation}
where $G_{OO}$ is the Greens function of the the operator
$O=\phi^*\phi$.  The OPE of two $O$ operators is easy to compute,
\begin{equation}
  -i \<T\, \phi^*\phi(t,\x)\, \phi^*\phi(0)\> = 
  \<\phi^*\phi\> [D(t,\x) + D(-t,-\x)] + \cdots, 
\end{equation}
where $D(t,\x)$ is the scalar propagator [Eq.~(\ref{scalar-prop})].
From this we find $\Im  G_{OO}(\omega,{\bf 0}) =
-4\pi\omega^{-1/2}\<\phi^*\phi\>$.  Thus we find the rate of energy
deposition,
\begin{equation}
  \frac{d\epsilon}{dt} = \frac{\alpha_0^2}{8\pi}\<\phi^*\phi\> \sqrt\omega\,.
\end{equation}

If $\alpha(t)$ has the form of a sudden pulse, i.e., is nonzero only
in a finite time interval, the total energy deposited into the system,
per unit volume, will be related to the Fourier transform
$\tilde\alpha(\omega)=\int\!dt\,e^{i\omega t}\alpha(t)$ by
\begin{equation}
  \epsilon = \frac{\<\phi^*\phi\>}{2\pi}\int_0^\infty\!\frac{d\omega}{2\pi}\,
  \sqrt\omega\, |\tilde\alpha(\omega)|^2.
\end{equation}

\section{Conclusion}

In this paper we have shown that the structure of a unitary Fermi gas,
at short distance and length scales, can be obtained from the operator
product expansion.  We computed the OPE of the density operator at two
points and derived an expression for the dynamic structure factor.
The same technique is applied to the problem of finding the energy
deposited into a unitary Fermi gas by a rapid, small oscillation of the
scattering length.

As mentioned above, the two-body contribution to $S(\q,\omega)$ vanishes 
for $\omega<q^2/4$: this is the two-body threshold.
If $\omega>q^2/6$, it is possible to transfer the energy and
momentum to three particles instead.  In general, when $n-1<x<n$, where $x$ is
the ``Bjorken $x$'' variable, 
the dominant
contribution to the dynamic structure factor $S(\omega,\q)$ in the large
$\omega$, fixed $x$ regime is due to the lowest dimensional operator in
the OPE that contains $n$ creation and $n$ annihilation operators.
The dimension of this operator is $2\Delta_n$, where $\Delta_n$ is the
ground state energy of a system of $n$ particles in a harmonic
potential.  Thus we find
\begin{equation}
  S(\q,\omega) \sim \<O_n^\+ O_n\> \frac{f_n(x)}{\omega^{\Delta_n-1/2}}\,,
\end{equation}
in the limit $\omega\to\infty$, $x=\textrm{fixed}$, and $n-1<x<n$ ($n\ge3$). 
 For example, the $n=3$
body contribution to the dynamic structure 
factor is down by $\omega^{-3.77272}$
since $\Delta_3\approx4.27272$.  The larger $n$ is, the faster the
contribution of $n$-body physics decreases with increasing momentum.

The operator product expansion cleanly separates ``hard''
(large-momentum and/or large-energy) physics from soft physics.  The
computation of the OPE coefficients involves computing diagrams in
vacuum, with a finite number of particles in the intermediate state.
Thus the OPE is a natural way of separating few-body from
many-body physics.  It application to cold atom physics should be
further investigated.

\acknowledgments
The authors thank G.~Baym, E.~Braaten, A.~L.~Fitzpatrick, E.~Katz,
Y.~Nishida, L.~Platter, S.~Shenker, and S.~Tan for discussions.  This
work is supported, in part, by DOE grant No.\ DE-FG02-00ER41132
and by University of Washington Royalty Research Fund grant No.\ 65-8195.
After this work was finished, we learned about
Ref.~\cite{Braaten:2010dv} which has some overlap with our calculation
of the RF response.  We also learned from M.~Randeria and E.~Taylor that
they considered
$S(\q,\omega)$ in the large-$\omega$, small-$q$ regime
and obtained the same $\<\phi^*\phi\>q^4/\omega^{7/2}$ behavior as 
in Eq.~(\ref{S-smallq}), but with 
a different numerical coefficient~\cite{Randeria}.

\end{document}